# Thermoelastic Equation of State of Boron Suboxide $B_6O$ up to 6 GPa and 2700 K: Simplified Anderson-Grüneisen Model and Thermodynamic Consistency

O.O. Kurakevych,[a] and V.L. Solozhenko [b,*]

[a] IMPMC, Sorbonne Universités – UPMC Univ Paris 6, Paris, France
[b] LSPM–CNRS, Université Paris Nord, Villetaneuse, France

*p-V-T* equation of state of superhard boron suboxide $B_6O$ has been measured up to 6 GPa and 2700 K using multianvil technique and synchrotron X-ray diffraction. To fit the experimental data, the theoretical *p-V-T* equation of state has been derived in approximation of the constant value of the Anderson-Grüneisen parameter $\delta_T$. The model includes bulk modulus $B_0 = 181$ GPa and its first pressure derivative $B_0' = 6$ at 300 K; two parameters describing thermal expansion at 0.1 MPa, i.e. $a = 1.4 \cdot 10^{-5}$ $K^{-1}$ and $b = 5 \cdot 10^{-9}$ $K^{-2}$, as well as $\delta_T = 6$. The good agreement between fitted and experimental isobars has been achieved to the absolute volume changes up to 5% as compared to volume at standard conditions, $V_0$. The fitted thermal expansion at 0.1 MPa is well consistent with the experimental data, as well as with ambient-pressure heat capacity $c_p$, bulk modulus $B_0$ and $\delta_T$ describing its evolution with volume and temperature. The fitted value of Grüneisen parameter $\gamma = 0.85$ is in agreement with previous empiric estimations for $B_6O$ and experimental values for other boron-rich solids.

**Introduction.**

Boron-oxygen system is very promising for both fundamental physics and chemistry, as well as for materials science [1]. Very unusual thermodynamic and kinetic properties of boron oxide $B_2O_3$ has been reported [2-5]; a number of intriguing compounds has been predicted [6-8], while to the present time only boron suboxide $B_6O$ has been synthesized and characterized [9-14]. Although the thermodynamics of boron at both ambient and high pressure has been recently studied quite thoroughly [15], the majority of boron-rich solids remains poorly investigated, especially, the phases that are synthesized at high pressure.

Boron suboxide $B_6O$ [9] is a superhard phase that is usually obtained by high pressure – high temperature synthesis [12,16]. $B_6O$ crystallizes in a *R-3m* space group and is able to form unusual multiply-twinned crystals with icosahedral habit [17]. The $B_{12}$ icosahedra form a distorted cubic close packing (ccp) with O-atoms connecting the icosahedra along the [111] rhombohedra direction [14]. $B_6O$ combines low density, high hardness and chemical inertness that are useful for shaping materials and high-wear application [18-20]; and unique thermal and electronic properties that may be useful for thermoelectric power generation [21]. At present the 300-K equation of state to 60 GPa [22], data on the heat capacity to 800 K [23] and some other thermodynamic properties [12,18] of $B_6O$ are established. Still, the experimental high pressure – high temperature thermodynamic data are lacking, which render difficult predictions and synthesis of new challenging materials, for example, in the B–N–O system at high pressures [24].

---

[*] e-mail: vladimir.solozhenko@univ-paris13.fr

In our previous study [18] we analyzed the available heat capacity data and fitted them to the adaptive pseudo-Debye model proposed by Holtzapfel (see [25] for details):

$$c_p = 3R\tau^3 \frac{4C_0 + 3C_1\tau + 2C_2\tau^2 + C_3\tau^3}{(C_0 + C_1\tau + C_2\tau^2 + C_3\tau^3)^2} \left[1 + A\frac{\tau^4}{(a+\tau)^3}\right] \qquad \text{Eq. 1}$$

where $\tau = T/\theta_h$; $\theta_h$ is Debye temperature in the high-temperature region; $C_1$, $C_2$ and $A$ are parameters to be fitted; $C_3 = 1$; $a$ characterizes non-harmonicity and was fixed to 1/8; R is gas constant. Parameter $C_0$ has been chosen as $C_0 = (5\,\theta_l^3)/(\pi^4\,\theta_h^3)$, with $\theta_l$ and $\theta_h$ Debye temperatures in the low- and high-temperature regions, respectively, in order to obtain these values directly as fitting parameters. The fitting gave $\theta_h = 1020$ K, $\theta_l = 1175$ K, $C_1 = -0.0865$, $C_2 = 1$, and $A = 0.125$. These values allowed establishing the best standard values of $c_p = 10.24$ J mol$^{-1}$ K$^{-1}$, $H°-H_0° = 1.147$ kJ mol$^{-1}$ and $S°-S_0° = 5.621$ J mol$^{-1}$ K$^{-1}$ [18]. Very recently, the new data on $B_6O$ heat capacity has been reported [26], and it would be important to improve our thermodynamic vision of $B_6O$ and check the challenging predictions [18].

The *p-V-T* equation of state is required to calculate the thermodynamic potentials at high pressures according to the method described elsewhere [11,27]. In the present work we have measured the thermal expansion of $B_6O$ in the 300-2700 K temperature range at pressures up to 6 GPa. Based on these data, the thermoelastic equation of state of $B_6O$ has been proposed.

**Experimental.**

Boron suboxide has been synthesized by the reaction of β-rhombohedral boron with $B_2O_3$ at 6 GPa and 2000 K (5-min heating) using a toroid-type high-pressure apparatus according to the procedure described in [13]. The recovered sample has reddish-brown color. X-ray powder diffraction (Seifert MZIII, Bragg-Brentano geometry, CuKα radiation) indicated that the sample contains only highly crystalline $B_6O$ phase. According to the full profile analysis of the diffraction pattern, the lattice parameters are $a = 5.385(2)$ Å, $c = 12.32(1)$ Å i.e. close to the values reported in [13]. Thus, the oxygen deficiency in as-synthesized $B_6O$ sample is negligible, because the dependence of lattice parameters on concentration of oxygen vacancies is significant, i.e. d$a$/d$x$ = 0.083 Å and d$c$/d$x$ = -0.117 Å (see the detailed sample characterization in [10]).

*In situ* high-pressure experiments up to 6 GPa were carried out using a multianvil X-ray system MAX80 at beamline F2.1, HASYLAB-DESY. The experimental setup has been described elsewhere [28]. Energy-dispersive data were collected on a Canberra solid state Ge-detector with fixed Bragg angle $2\theta$ = 9.110(3)° using a white beam collimated down to 60×100 μm$^2$ (vertical by horizontal) and the detector optics with $2\theta$ acceptance angle of 0.005°, which ensures a high resolution of the observed diffraction patterns. The detector was calibrated using the $K_\alpha$ and $K_\beta$ fluorescence lines of Cu, Rb, Mo, Ag, Ba, and Tb. To decrease the deviatoric stress that was generated during "cold" compression and thus attain quasi-hydrostatic pressure condition during equation-of-state measurements, the samples were pre-annealed at 800 K and a given pressure for 10 min. The sample pressure was determined from the $c$ lattice parameter of highly-ordered hBN using the corresponding equation of state reported in [29] and/or by the equation of state of $B_6O$ [22]. The temperature was measured by W3%Re–W25%Re thermocouple (without correction for the pressure effect on the thermocouple emf) and estimated using power–temperature calibration curve at temperatures above the thermocouple melting. Additionally, CsCl melting curve [30] was used as a control in some experiments.

**Results and Discussion.**

*Experimental data.*

The energy-dispersive diffraction patterns of $B_6O$ show three well-defined lines, i.e. *003*, *104* and *021*, that were used for the estimation of lattice parameters and unit cell volumes at each experimental point (Fig. 1a). The temperature dependences of the relative volume $V/V_0$ ($V_0$ corresponds to 300 K and ambient pressure) at different pressures are shown in Fig. 2. In the 300-2500 K range the dependences are very close to linear (the points at 300 K correspond to the equation of state measured in [22]). The slopes, however, remarkably depend on pressure.

At temperatures above 2500 K, the strong weakening of *003* and surrounding (*101* and *012*) reflections has been occasionally observed (Fig. 1a). Similar patterns have been previously reported in the case of the B–N system [31,32], but they do not belong to any phase of the "$B_6O$-type" or to more general "α-boron related" structures [33,34]. The appearance of a new phase may be the reason why some point at highest temperatures deviate from the theoretical equations of state. However, some pressure drop can be also a reason for that. Here for processing of the equation-of-state data we have used only characteristic diffraction patterns of $B_6O$.

*Simplified Anderson-Grüneisen model.*

For the fit of the thermoelastic data, we have used the Anderson-Grüneisen model [35] combined with the results of [36,67] (see [38] for details and physical reasoning), i.e.

$$\alpha(p,T) = \alpha(0,T)\left[\frac{V(p,T)}{V(0,T)}\right]^{\delta_T} \quad \text{Eq. 2}$$

or, in the terms of bulk modulus,

$$B(p,T) = B(p,300)\left[\frac{V(p,T)}{V(p,300)}\right]^{-\delta_T} \quad \text{Eq. 3}$$

If $\delta_T$ for both thermal expansion and bulk modulus is constant, Eq. 2 can be easily transformed (using the definition of thermal expansion) into

$$\int_{T_1}^{T_2} d[V(p,T)]^{-\delta_T} = \int_{T_1}^{T_2} d[V(0,T)]^{-\delta_T}, \quad \text{Eq. 4}$$

and, therefore, we obtain

$$V(p,T) = \left[V(0,T)^{-\delta_T} + V(p,300)^{-\delta_T} - V(0,300)^{-\delta_T}\right]^{-1/\delta_T}. \quad \text{Eq. 5}$$

To a good approximation the above relationships can be used when the volume change due to compression/thermal dilatation does not exceed 5-10%, i.e. when Eqs. 2 & 3 are valid [39].

During fitting procedure, the $V(0,T)$ dependence [40,41] has been suggested to follow equation

$$V(0,T) = V(0,300)\ [1+a\ (T-300)+b\ (T-300)^2], \quad \text{Eq. 6}$$

while the $V(p,300)$ dependence has been fixed to the 300-K equation of state of $B_6O$ reported in [22].

Fitting all available experimental $p$-$V$-$T$ data to Eq. 5 leads to $\delta_T = 6$, $a = 1.4 \cdot 10^{-5}$ K$^{-1}$, $b = 5 \cdot 10^{-9}$ K$^{-2}$, while $V(p,300)$ is defined by Vinet equation of state [42]:

$$p = 3B_0(V/V_0)^{-2/3}\left[1-(V/V_0)^{1/3}\right]e^{1.5\ (B'_0-1)\left[1-(V/V_0)^{1/3}\right]} \quad \text{Eq. 7}$$

with $B_0 = 181$ GPa and $B_0' = 6$ [22].

*Thermal expansion coefficient and bulk modulus at high pressure and high temperature.*

The pressure dependence of mean thermal expansion and temperature dependence of bulk modulus are shown in Figs. 3a and 3b, respectively. The experimental data (symbols) well fit the curves calculated using the model represented by equations (1) and (2). The thermal expansion at 0.1 MPa and 2000 K (the mean value of the 1500-2500 K region) should be 3.4 K$^{-1}$ using *a*- and *b*-coefficients described above. This value is in a good agreement with 3.6 K$^{-1}$, obtained by exponential extrapolation of our experimental data down to 0.1 MPa (Fig. 3a), which additionally validates our thermoelastic equation of state. The mean high-temperature $\alpha$ as a function of pressure well follows the exponential dependence:

$$<\alpha> = 3.567 \cdot 10^{-5} \cdot e^{-0.09546\ p}. \qquad \text{Eq. 8}$$

Thus, B$_6$O volume at given *p-T* conditions may also be calculated using the following approximate equation:

$$V(T,p) = V(300,p) \cdot \exp\left(3.567 \cdot 10^{-5} \cdot e^{-0.09546\ p} \cdot (T-300)\right). \qquad \text{Eq. 9}$$

This equation allows us to make a good estimation for high-temperature behavior of bulk modulus, i.e.

$$\left(\frac{\partial \alpha}{\partial p}\right)_T = \left(\frac{\partial}{\partial p}\left(\frac{\partial \ln V}{\partial T}\right)_p\right)_T = \left(\frac{\partial}{\partial T}\left(\frac{\partial \ln V}{\partial p}\right)_T\right)_p = \left(\frac{\partial}{\partial T}\left(\frac{1}{B}\right)_T\right)_p \qquad \text{Eq. 10}$$

and, after integration over the temperature,

$$\int_{300}^{T}\left(\frac{\partial \alpha}{\partial p}\right)_T dT = \frac{1}{B_0(T)} - \frac{1}{B_0(300K)} \qquad \text{Eq. 11}$$

and the temperature dependence of a bulk modulus is expressed by equation

$$B_0(T) = \frac{B_0(300K)}{1 - B_0(300K)\int_{300}^{T}\left(\frac{\partial \alpha}{\partial p}\right)_T dT}. \qquad \text{Eq. 12}$$

The pressure derivative of the thermal expansion coefficient does not depend on temperature (at least, for mean values at high temperatures); so the zero-pressure bulk modulus of B$_6$O should change with temperature according to equation

$$B_0(T) \cong \frac{B_0(300K)}{1 + B_0(300K) \cdot 3.403 \cdot 10^{-6}(T-300)}. \qquad \text{Eq. 13}$$

The temperature dependence of $B_0$ is illustrated in Fig. 3. Two values of $B_0(300)$, 181 GPa [22] and 217 GPa (the value suggested by McMillan, private communication), have been used for the construction of the theoretical curves (Eq. 13). The bulk moduli were established at high (1000-2000 K) temperatures by fitting the $p$-$V$ data to the Murnaghan equation [43] at a given temperature ($B_0'$ fixed to 6 and independent of temperature). When the Vinet equation was used, the difference with the Murnaghan equation is less than 0.5%, at least up to 10 GPa. Thus, our high pressure – high temperature data well agree with room-temperature equation of state reported in [22], and can serve an independent confirmation of the 300-K bulk modulus value of $B_0 = 181$ GPa.

*Thermodynamic consistency of equation of state.*

The thermodynamic consistency of thus established $p$-$V$-$T$ equation of state and heat capacity $c_p$ experimental data [18,23,26] has been verified. The relationship between thermal expansion and heat capacity $c_v$ may be expressed as

$$\alpha = \frac{\gamma \cdot c_v}{B \cdot V}, \qquad \text{Eq. 14}$$

where $\gamma$ is Grüneisen parameter and $B$ is isothermal bulk modulus. Heat capacity $c_p$ is given by the following equations:

$$c_p = c_v + \alpha^2 BTV \quad \text{and} \quad c_p = \alpha \frac{BV}{\gamma} + \alpha^2 BTV, \qquad \text{Eq. 15}$$

and, hence, thermal expansion can be expressed as

$$\alpha = \frac{1}{2\gamma T}\left(\sqrt{1 + \frac{4 c_p \gamma^2 T}{BV}} - 1\right) \qquad \text{Eq. 16}$$

Now, we can integrate Eq. 16 and obtain the integral functional equation in respect to $V(T)$:

$$\ln\frac{V(T)}{V_0} = \int_{T_0}^{T} \frac{1}{2\gamma\tau}\left(\sqrt{1 + \frac{4 c_p(\tau)\,\gamma^2 \tau\, V_0^{\delta_T}}{B_0[V(\tau)]^{\delta_T+1}}} - 1\right) d\tau \qquad \text{Eq. 17}$$

In order to solve Eq. 17 in respect to $V(T)$, one can easily use the iteration method

$$V_{i+1}(T) = f(V_i(T), \gamma). \qquad \text{Eq. 18}$$

where

$$f(V(T),\gamma) = V_0 \exp\left\{\int_{T_0}^{T} \frac{1}{2\gamma\tau}\left(\sqrt{1 + \frac{4c_p(\tau)\,\gamma^2\tau\,V_0^{\delta_T}}{B_0[V(\tau)]^{\delta_T+1}}} - 1\right)d\tau\right\}$$ Eq. 19

Parameter $\gamma$ in Eqs. 17-19 remains unknown and should be fitted, contrary to $\delta_T$, $V_0$ at 300 K and 0.1 MPa, and $c_p(T)$. As first approximation of the $V_1(T)$ function we took the dependence defined by Eq. 6. Five iterations usually give a stable solution of Eq. 17, i.e. $V(T) \cong V_5(T) \pm 0.1\%$.

Fig. 4a shows the heat capacity $c_p$ of $B_6O$ as a function of temperature. Three experimental sets of data are presented: one corresponding to the low-temperature measurements [23], one high-temperature set up to 780 K [23] and very recent data up to 1250 K [26]. The latter gives higher heat capacity values in the high-temperature range, and the fitting parameters for Eq. 1 are $\theta_h = 1075$ K, $\theta_l = 1275$ K, $C_1 = -0.05$, $C_2 = 0.75$, and $A = 0.10$. These new heat capacity data show the best agreement with experimental $V(T)$ dependence [40,41] (Fig. 4b). The Grüneisen parameter $\gamma$ can be estimated as 0.85 (high-temperature limit), in a reasonable agreement with previous empirical estimations of ~0.75(15) (mean value over the 50-1500 K range) [40,41].

**Conclusions.**

Thermal expansion of $B_6O$ has been studied in the 300-2700 K temperature range at pressures from 2 to 6 GPa. Based on these data, thermoelastic equation of state has been constructed which is well consistent with independent experimental data on 300-K equation of state and thermal expansion at 0.1 MPa. To fit the experimental data, the theoretical $p$-$V$-$T$ equation of state of $B_6O$ has been proposed. The model implies the constant value of Anderson-Grüneisen parameter $\delta_T = 6$; bulk modulus $B_0 = 181$ GPa and its first pressure derivative $B_0' = 6$ at 300 K; and two parameters describing the thermal expansion at 0.1 MPa, i.e. $a = 1.4 \cdot 10^{-5}$ K$^{-1}$ and $b = 5 \cdot 10^{-9}$ K$^{-2}$.

**Acknowledgements.** *In situ* experiments at HASYLAB-DESY have been performed during beamtime allocated to Projects DESY-D-I-20080149 EC and DESY-D-I-20120445 EC and received funding from the European Community's Seventh Framework Programme (FP7/2007-2013) under grant agreement n° 226716. This work was financially supported by the Agence Nationale de la Recherche (grant ANR-2011-BS08-018).

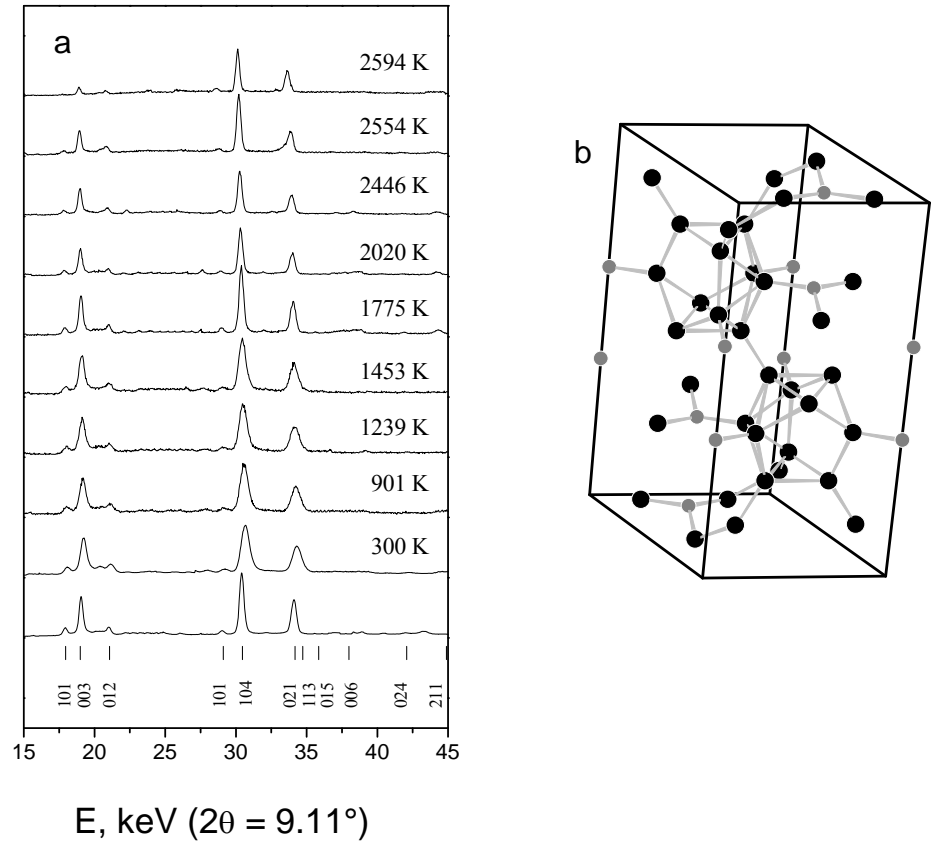

**Figure 1.** (*a*) Energy-dispersive diffraction patterns of $B_6O$ taken *in situ* at 4.3 GPa upon heating to 2600 K (the bottom pattern is taken at ambient conditions). (*b*) Crystal structure of $B_6O$. Black and grey balls represent boron and oxygen atoms, respectively.

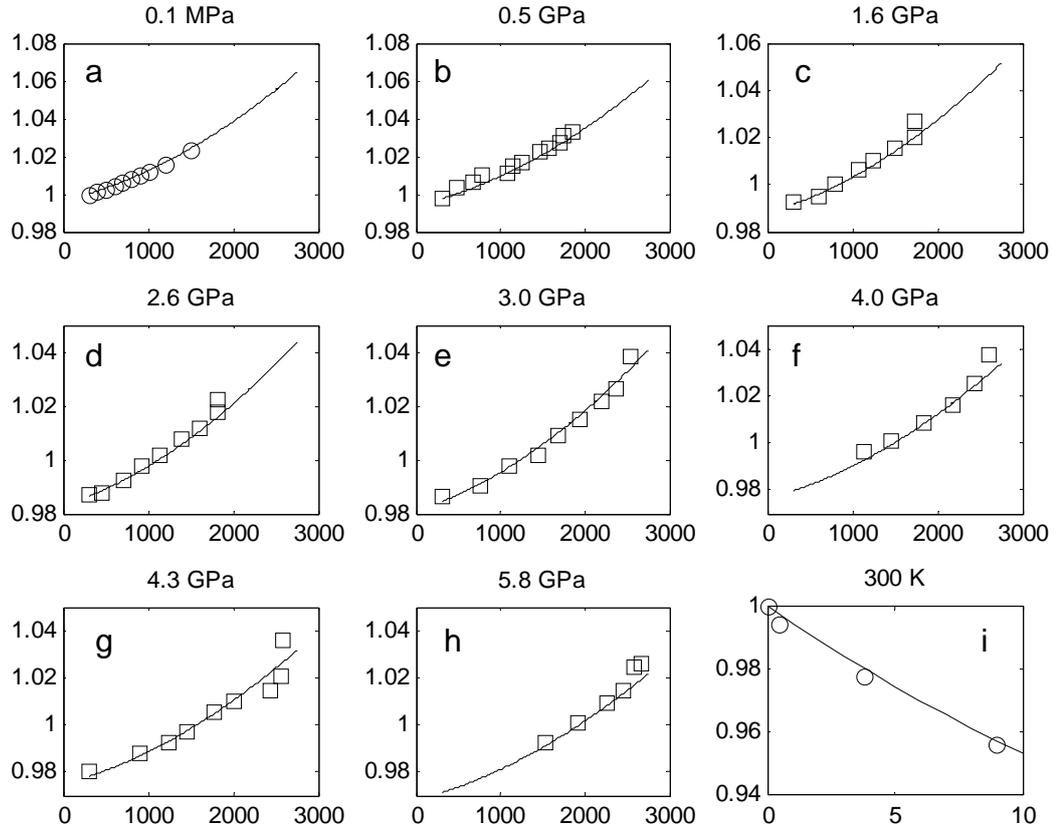

**Figure 2.** (*a - h*) Temperature dependence of relative volume $V/V_0$ of $B_6O$ ($V_0$ taken at 300 K and ambient pressure). The plot coordinates are $x - T$ (K) and $y - V/V_0$. (*i*) Pressure dependence of relative volume $V/V_0$ of $B_6O$ at 300 K. The plot coordinates are $x - p$ (GPa) and $y - V/V_0$.

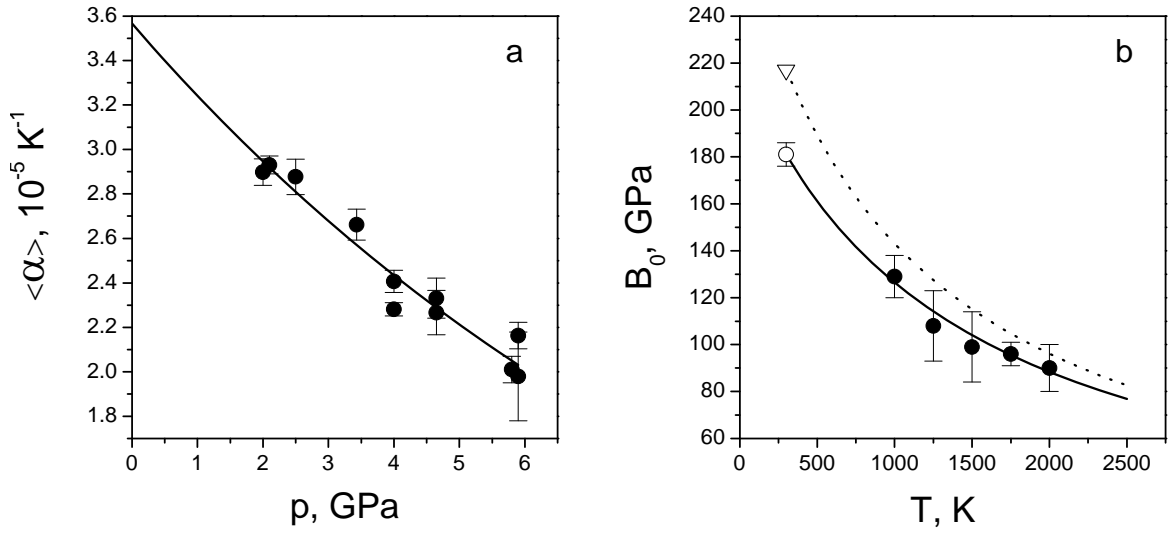

**Figure 3.** Experimental data on (**a**) pressure dependence of the mean high-temperature coefficient of thermal expansion of $B_6O$ (solid line shows the exponential fit) and (**b**) temperature dependence of the bulk modulus of $B_6O$ (solid line corresponds to the theoretical line with $B_0 = 181$ GPa, while dotted line, to $B_0 = 217$ GPa at 300 K). Solid symbols represent our results, open symbols are literature data [22].

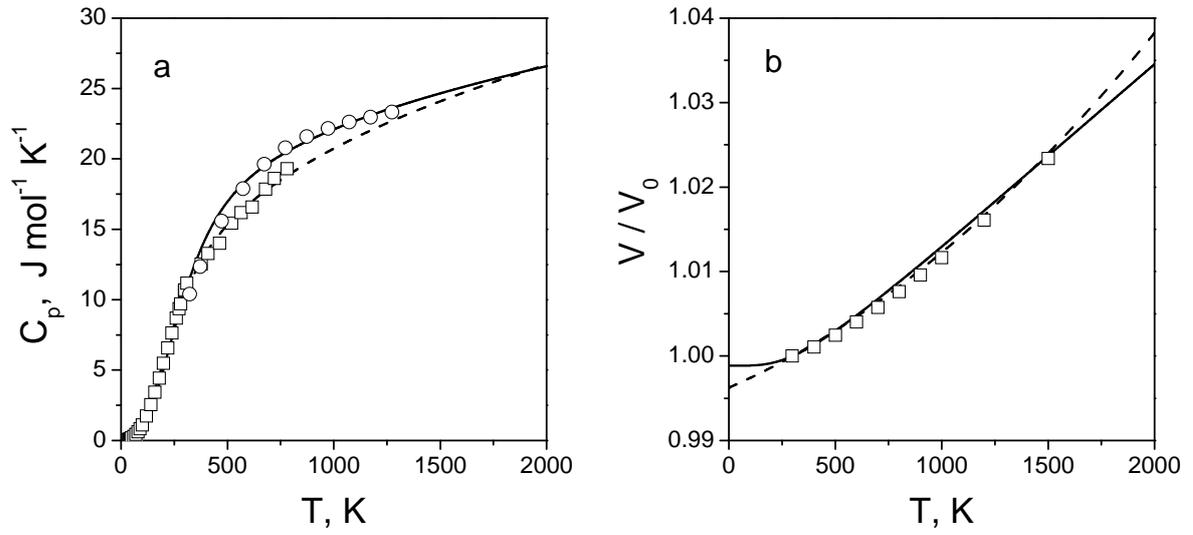

**Figure 4.** Experimental data on (*a*) heat capacity $c_p$ of $B_6O$ as a function of temperature (squares from [23], circles from [26], lines show the results of fitting to Eq. 1); and (*b)* thermal expansion of $B_6O$ at 0.1 MPa (squares are experimental data [41], dashed curve corresponds to the square-polynomial fitting (Eq. 6), while solid one – to the numerical solution of Eq. 17 with Grüneisen coefficient $\gamma = 0.85$).